\RequirePackage{lineno}
\documentclass[aps,prl,twocolumn,superscriptaddress]{revtex4}
\usepackage{color,graphicx,calc,url,amsfonts,amsmath,amssymb}
\usepackage{datetime}
\usepackage[dvipdfm]{hyperref}
\providecommand{\U}[1]{\protect\rule{.1in}{.1in}}

\usepackage{lineno}
\begin{document}
\title{Neutron optical beam splitter from holographically structured nanoparticle-polymer composites}
\author{M. Fally}
\affiliation{Faculty of Physics, Nonlinear Physics, University of Vienna, Boltzmanngasse 5, A-1090 Wien, Austria}
\homepage{http://fun.univie.ac.at}
\email{martin.fally@univie.ac.at}
\author{J. Klepp}
\affiliation{Faculty of Physics, Nonlinear Physics, University of Vienna, Boltzmanngasse 5, A-1090 Wien, Austria}
\author{Y. Tomita}
\author{T. Nakamura}
\affiliation{Department of Electronics Engineering, University of Electro-Communications, 1-5-1 Chofugaoka, Chofu, Tokyo 182, Japan}
\author{C. Pruner}
\affiliation{Department of Materials Science and Physics, University of Salzburg, A-5020 Salzburg, Austria}
\author{M. A. Ellabban}
\affiliation{Physics Department, Faculty of Science, Tanta University, Tanta 31527, Egypt}
\author{R. A. Rupp}
\affiliation{Faculty of Physics, Nonlinear Physics, University of Vienna, Boltzmanngasse 5, A-1090 Wien, Austria}
\affiliation{Nankai University, TEDA Applied Physics Faculty, Tianjin 300457, P.R. China}
\author{M. Bichler}
\affiliation{Faculty of Physics, Nonlinear Physics, University of Vienna, Boltzmanngasse 5, A-1090 Wien, Austria}
\author{I. Dreven\v{s}ek Olenik}
\affiliation{Faculty of Mathematics and Physics, University of Ljubljana, Jadranska 19, SI 1001 Ljubljana, Slovenia}
\affiliation{ J. Stefan Institute, Jamova 39, SI 1001 Ljubljana, Slovenia}
\author{J. Kohlbrecher}
\affiliation{Laboratory for Neutron Scattering, ETH Zurich \& Paul Scherrer Institut, 5232 Villigen PSI, Switzerland}
\author{H. Eckerlebe}
\affiliation{GKSS Forschungszentrum, D-21502 Geesthacht, Germany}
\author{H. Lemmel}
\affiliation{Institut Laue Langevin, Bo\^{i}te Postale 156, F-38042 Grenoble  Cedex 9, France}
\author{H. Rauch}
\affiliation{Institut Laue Langevin, Bo\^{i}te Postale 156, F-38042 Grenoble Cedex 9, France}
\affiliation{Atominstitut, Technische Universit\"at Wien, Stadionallee 2, 1020 Wien, Austria}
\date{\today, \currenttime, \jobname}
\newcommand{\zro}{\mbox{$\rm ZrO_2$}}
\newcommand{\sio}{\mbox{$\rm SiO_2$}}
\newcommand{\cwidth}{\columnwidth}
\begin{abstract}
We report a breakthrough in the search for versatile diffractive elements for cold neutrons. Nanoparticles are spatially arranged by holographical means in a photopolymer. These grating structures  show remarkably efficient diffraction of cold neutrons up to about 50\% for effective thicknesses of only 200 micron. They open up a profound perspective for next generation neutron-optical devices with the capability to tune or modulate the neutron diffraction efficiency.
\end{abstract}
\maketitle

Neutron-optical phenomena arise from coherent elastic scattering. They are described by the wave equation for neutrons. The essential material property is the neutron-optical potential or, equivalently, the neutron refractive index at wavelength $\lambda$ \cite{Sears-89}.
The basic diffractive element is a (one-dimensional) sinusoidal grating characterized by a refractive index $n(x)=n_0+n_1 \cos(K x)$ which is periodically modulated with the spatial frequency $K$ and amplitude $n_1$. Depending on its diffraction efficiency, it can be used as beam splitter, mirror or monochromator, and it can be altered and arranged to form more complex devices for imaging, spectroscopy, and in particular, cold neutron interferometry \cite{Schellhorn-phb97,Pfeiffer-prl06,Pruner-nima06}.
Efficient neutron optical devices are vital for any neutron experiment,  either as part of the instrumentation (e.g. monochromators, guides, calibration standards) or as a tool to investigate fundamental physical questions (e.g. interferometers).
 Neutron optics - and in particular interferometry with thermal neutrons using perfect crystals as diffractive elements  - has reached a satisfying level giving rise to an appealing insight on quantum mechanical problems \cite{Rauch-00}.  As the difference of the refractive index for neutrons from unity depends quadratically on the wavelength the latter should be further increased to maximize the neutron-optical potential. On the other hand the neutron flux decreases dramatically at low energies. On top of that interferometry experiments with very cold neutrons, so far, employed ruled or blazed gratings as thin diffraction elements. Consequently various diffraction orders are excited at the same time resulting in even lower measuring signals \cite{Zouw-nima00}. Therefore, cold neutrons offer a fair trade-off between flux and magnitude of the neutron-optical potential. Moreover, small angle neutron scattering (SANS) beamlines are workhorses at neutron facilities due to the interdisciplinary  interests of biologists or medical scientists when exploring complex large-scale soft matter, see e.g. \cite{Teixeira-cp08}. A few attempts have been conducted to manufacture diffractive elements for cold neutrons.  However, the results on the efficiency were rather discouraging \cite{Rupp-prl90,Funahashi-pra96,Schellhorn-phb97,Fally-prl06}.

In this Letter we introduce holographically patterned nanoparticle-polymer composite systems as efficient and extremely flexible diffractive elements for cold neutrons.

Holographic production of neutron-optical elements can be performed in materials where the neutron refractive-index can be changed by photon irradiation. This is the so-called photo-neutronrefractive effect (PNRE) \cite{Rupp-prl90,Fally-apb02}. Upon illumination with a spatial light pattern a corresponding refractive-index modulation $\Delta n=(\lambda^2/2\pi)\Delta{\cal B}$ occurs that is proportional to the local change
 $\Delta {\cal B}$ of the coherent scattering length density ${\cal B}$. In the simplest case ${\cal B}=b_c\rho$ is the product of the coherent scattering length $b_c$ and the number density of atoms $\rho$. The PNRE may result from a photo-induced modulation of either of them.

Up to now only single-component systems were investigated and no suitable light sensitive materials have been found so far. Either fundamental problems on the neutron-optical or on the light-optical side turned out to be too difficult to overcome. This has recently led us to turn the attention to multicomponent systems. Recently, a successful attempt was made by employing holographic polymer dispersed liquid crystals \cite{Fally-prl06}.
The  figure of merit - the coherent scattering length density modulation - is strong ($\Delta{\cal B}\approx 10~\mu$m$^{-2}$) in these materials. However, the disadvantage is its restricted neutron diffraction efficiency due to a limited grating thickness $d$ of about only $30~ \mu$m. 
The reason for this is a strong diffuse scattering and/or coherent light-induced scattering (holographic scattering \cite{Ellabban-apl05}) originating from the anisotropy of the liquid crystals component which destroys the incident light pattern when propagating through the sample. 
For low efficiencies of less than a few percent the relation to the coherent scattering length density modulation is independent of the diffraction regime and reads $\eta_1= \nu^2$ with the normalized coupling length $\nu=\lambda\Delta{\cal B}d/2$.

Preparation and handling of various nanoparticle-polymer  composites nowadays have reached a pretty mature state \cite{Zhao-nmat09}. In particular, the groups of {\it Tomita et al. } \cite{Suzuki-apl02,Tomita-ol05,Suzuki-oex06,Nakamura-joa09} and {\it Stumpe et al. } \cite{Sakhno-nt07,Sakhno-joa09} have intensively investigated inorganic nanoparticles (\sio~and \zro) embedded in a photopolymer matrix.
They serve as proof of principle for the essential advantages offered by this class of materials for neutron optics over previously used pure photopolymer systems \cite{Rupp-prl90,Rupp-p97,Fally-apb02} :
the refractive-index modulation can be tuned by the type of nanoparticles, their size and volume ratio. 
%
An additional benefit of including nanoparticles constitutes the increased mechanical stability because shrinkage, which is typical for polymerization processes, is strongly reduced \cite{Suzuki-apl02}. Stability is a decisive advantage when it comes to the construction of cold neutron interferometers since it permits to produce thicker gratings with mirror-like reflection. For light-optical purposes it has indeed already been shown that grating thicknesses up to 0.2 mm can be achieved with only moderate disturbance of the interference pattern by holographic scattering \cite{Suzuki-ao04,Suzuki-ao07}.

\sio~(\zro) nanoparticles having an average core diameter of approximately 13 (3) nm were prepared by liquid-phase synthesis and dissolved in a methyl isobutyl ketone (toluene) solution. The \sio~(\zro) sol was dispersed to methacrylate monomer [2-methyl-acrylic acid 2-[4-[2-(2-methyl-acryloyloxy)-ethylsulfanylmethyl]-benzylsulfanyl]-ethyl ester,
(acrylate monomer, 2-propenoic acid,(octahydro-4,7-methano-1H-indene-2,5-diyl)bis(methylene)-ester),) whose
refractive indices are 1.59 (1.53) in the solid phase at 589 nm. At this wavelength the refractive index of the surface-treated \sio~(\zro) nanoparticles is 1.46 (1.71).

We used 1wt.\% titanocene (Irgacure784, Ciba) as radical photoinitiator enabling the monomer to photopolymerize at wavelengths shorter than 550 nm. The mixed syrup was cast on a spacer loaded glass plate and was dried before covering it with another glass plate to obtain film samples for light-optical characterization as well as neutron-optical measurements.
Both sample types, the one incorporating \sio~ nanoparticles (referred to as samples \textbf{S} below) and the one with \zro~ nanoparticles (referred to as  sample \textbf{Z} below), contained the same nanoparticle concentration of 20 vol.\%. Further details on our sample preparation technique can be found in Refs. \cite{Suzuki-ao07,Suzuki-oex06}. A conventional two-beam interference setup was used for recording an unslanted transmission grating employing two mutually coherent and $s$-polarized laser beams of equal intensities at $\lambda_L=532$ nm.
A series of such \textbf{S}-samples with different thicknesses $d_0$ and grating spacings $\Lambda$ and a \textbf{Z}-sample  were prepared.
In what follows we focus our investigations on only three of them: \textbf{Z0}  and \textbf{S0}  ($d_0=50~\mu$m, $\Lambda=0.5~\mu$m) as well as \textbf{S7} ($d_0=100~\mu$m, $\Lambda=0.5~\mu$m).

In order to relate the properties of the actual gratings to those previously investigated \cite{Suzuki-ao04,Suzuki-oex06} we first characterized them by light optical diffraction experiments.
They reveal two major common features for the gratings employing different nanoparticle species:
i) diffraction occurs clearly in the two-wave coupling regime (Bragg regime, i.e., only two diffracted beams propagate simultaneously) and ii) the gratings are obviously nonsinusoidal for both nanoparticle species which is indicated by the existence of higher diffraction orders. The latter is due to a nonlinear phase separation mechanism between nanoparticles and polymerizing monomer during holographic recording \cite{Tomita-ol05}. Further, we noticed that i) the refractive-index modulation for light is by about a factor of two larger for the sample \textbf{Z0} as compared to the corresponding sample \textbf{S0}, and ii)  a smaller grating spacing leads to a reduction of the refractive-index modulation \cite{Suzuki-apl02,Suzuki-oex06,Suzuki-ao04,Sakhno-joa09}.

The SANS experiments were conducted using cold neutrons at the spallation source SINQ, SANS-1 (Paul-Scherrer Institute) and the Geesthacht Neutron Facility, SANS-2 instrument. Measurements were performed at various mean neutron wavelengths between $\lambda$=1.16 nm and 2 nm having a distribution of about $\Delta\lambda/\lambda$=10\%. For the benefit of sufficient lateral coherence, the angular spread was restricted to less than 1 mrad by using two narrow slits and full collimation distances. The diffracted intensities were measured at a sample-detector distance of about 20 m  by a two-dimensional detector.

The experiments at $\lambda=1.16$ nm yield first order diffraction efficiencies of only $\eta_{1}=3$\% and $\eta_{1}=1$\% for the samples \textbf{S0} and  \textbf{Z0}, respectively. Evaluating the coherent scattering length density modulations, we find $\Delta{\cal B}=6.0~ \mu$m$^{-2}$ and $\Delta{\cal B}=3.1 ~\mu$m$^{-2}$. The obtained values are in the range of  those found in holographic polymer dispersed liquid crystals  with hitherto strongest neutron diffraction \cite{Fally-prl06}. However, nanoparticle-polymer composites do not share the drawbacks of those materials such as the holographic light scattering obstacle \cite{Ellabban-apl05}.
Moreover, neutron diffraction is more efficient for the \textbf{S} than for the \textbf{Z} sample which is contrary to the light optic result. Therefore, we focussed our further investigations only on the \sio~nanoparticle-composites having a different grating spacing ($\Lambda=0.5~\mu$m) and thickness.
From the light optical experiments we expect that a smaller grating spacing also reduces the coherent scattering length density modulation for neutrons, an assumption that is completely confirmed by measuring the diffraction signal from \textbf{S7}: a diffraction efficiency of $\eta_{1}=6$\% at a wavelength of $1.7$ nm yields a value of $\Delta {\cal B}=3.2 ~\mu$m$^{-2}$.

\begin{figure}[h]
 \includegraphics[width=\cwidth-1cm]{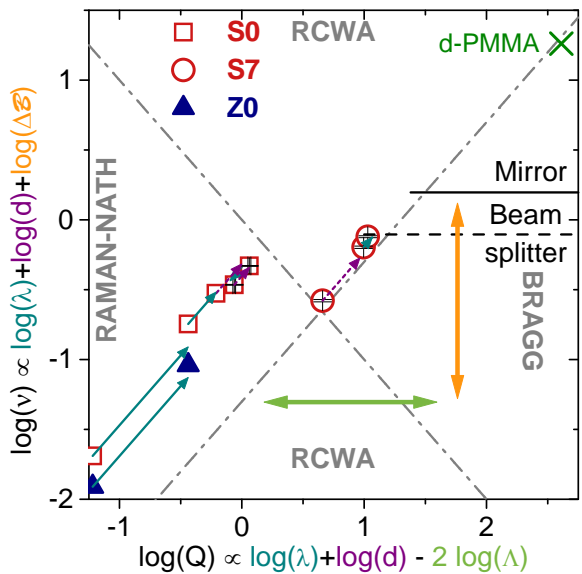}
\caption{\label{fig:4}Diffraction regimes separated by grey dashed-dotted lines and experimental results for \textbf{S0} (red  squares), \textbf{Z0}  (blue triangles) and \textbf{S7} (red circles) measured with thermal and cold neutrons. For comparison d-PMMA (d=3500 $\mu$m, $\Lambda=381$ nm, $\lambda=2.7$ nm) is shown. Dashed or solid diagonal arrows indicate a change in $d$ or $\lambda$.}
    \end{figure}

To facilitate the search for a route to create beam splitters or mirrors  the parameter space $(\log{(\nu)},\log{( Q)})$ is divided into three diffraction regimes, determined by the equalities $Q\nu=1 \wedge Q=20\nu$ as shown in Fig. \ref{fig:4} \cite{Gaylord-ao81}.  
Here,  the coupling length $\nu$ and $Q=2\pi\lambda d/\Lambda^2$ are the decisive parameters \cite{Extermann-hpa36,Gaylord-ao81}.
 Each of these regimes is governed by a particular diffraction theory: the two-wave coupling regime (Bragg) described by Kogelnik's theory \cite{Kogelnik-bell69}, the multiwave coupling regime neglecting dephasing (Raman-Nath theory \cite{Raman-piasa35}), and the regime to be evaluated according to the RCWA \cite{Moharam-josa81}.
Starting from the $(\nu,Q)$-values for the samples \textbf{Z0, S0} we see that even at the larger wavelengths we reside in the Raman-Nath diffraction regime. Nevertheless, the good news is that from these first experiments we could deduce a roadmap to produce beam splitters operating in the two-wave coupling regime. To move from the Raman-Nath to the Bragg diffraction regime just the thickness or wavelength (diagonal motion) must be increased and - most importantly -  the grating spacing $\Lambda$ decreased (horizontal motion). In Fig. \ref{fig:4} the experimentally obtained $(\nu,Q)$-values are shown together with  arrows indicating the type of parameter that was changed.
A further possible path (vertical motion), i.e., in- or decreasing ${\Delta\cal B}$, can be easily realized by simply altering the volume ratio or the species of the nanoparticles.  We would like to mention that measurements at thermal neutron wavelengths for \textbf{S0} and \textbf{Z0} ($\lambda=0.19$ nm) unexpectedly yield 25\% lower values of ${\Delta\cal B}$ which is far out of the experimental error and not yet understood.

\begin{figure}[h]
\includegraphics[width=\cwidth]{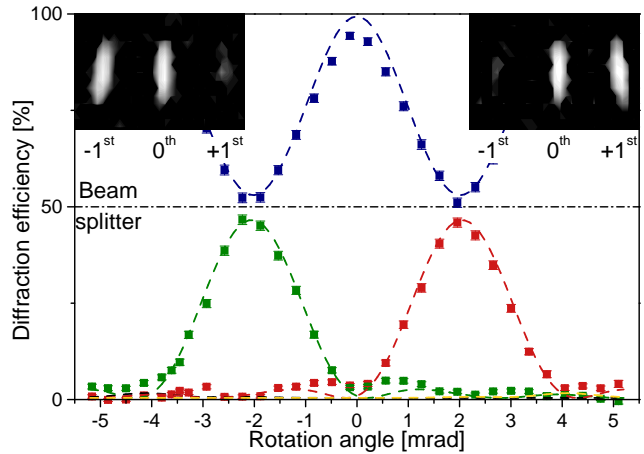}
\caption{\label{fig:3}Angular dependence of the $0^{\rm th}$ and $\pm 1^{\rm st}$ order diffraction efficiencies for cold neutrons at  $\lambda=2$ nm for gratings assembled from a \sio~nanoparticle-polymer composite with a spacing of $\Lambda=500$ nm and an effective thickness of $d=203 \mu$m (\textbf{S7}). Lines are a simultaneous RCWA fit to the data. Inset: Detector matrix (grey scale of logarithmic box counts at Bragg positions)}
    \end{figure}

To demonstrate the high potential of these types of gratings  we carefully chose the tunable parameters $d$ and $\lambda$  using sample \textbf{S7} with $\Lambda=0.5~ \mu$m. 
The results are  shown in Fig. \ref{fig:3}.
Here, the measured angular dependence of the efficiencies for $0^{\rm th}$ and $\pm 1^{st}$ diffraction orders of sample \textbf{S7} is depicted. It is obvious that the grating serves nearly as a perfect 50:50 beam splitter! Neutron diffraction in this case occurs essentially  in the two-wave coupling (Bragg) regime.  Then the corresponding coupling length $\nu$ is given by $\nu=\arcsin{(\sqrt{\eta_{1}})}$  for the $1^{\rm st}$-order diffraction \cite{Kogelnik-bell69}. For an accurate evaluation taking into account also the small contributions from multi-wave coupling we employed a fit based on the rigorous coupled wave analysis (RCWA) \cite{Moharam-josa81} included in Fig. \ref{fig:3}.

Switching back to Fig. \ref{fig:4}, the three experimental points for \textbf{S7} correspond to the parameter pairs ($\lambda=1.7$ nm, $d=d_0$), ( $\lambda=1.7$ nm, $d\sim 2 d_0$) and ($\lambda=2$ nm, $d\sim 2d_0$). These latter values allowed us to realize a beam splitter by employing  (1) a reduced  grating spacing as compared to \textbf{S0,Z0},  (2) an increased wavelength of $\lambda=2$ nm as well as (3) an  increased  effective thickness $d=d_0/\cos(\zeta)$ by an inclination of the sample for an angle $\zeta$ around the grating vector \cite{Somenkov-ssc78}.

Let us make a final note on why it is suboptimal to try the apparently simplest solution for producing a mirror, namely to use, e.g.,  deuterated poly methylmethacrylate (d-PMMA) and to increase the grating thickness to a few millimeters like done in a previous study \cite{Pruner-nima06}. The experimentally obtained value is also shown in Fig. \ref{fig:4}.
The disadvantage of such geometrically thick gratings is their extremely high angular selectivity. In this case the rocking curve becomes a very rapidly oscillating function of the Off-Bragg parameter. Since the  angular width of the rocking curve is usually much narrower than the angular spread of the beam, this results in averaging over the 'Pendell\"osung'-fringes \cite{Shull-prl68,Sears-89} with a mean reflectivity that cannot exceed 50\%  - an issue  that is well-known for standard perfect crystal silicon lattices used in thermal neutron interferometry \cite{Rauch-00}.

In future we could even take advantage of the spin dependent interaction potential between nanoparticles carrying a nuclear spin and polarized neutrons, e.g., superparamagnetic or multiferroic nanoparticles. This possibility opens up new perspectives of switchable gratings by external fields.

In summary we have realized a beam splitter for cold neutrons on the basis of nanoparticle-polymer composites, i.e., a first order diffraction efficiency of about 50\% for a  sample with 200 micron effective thickness. By reducing the grating spacing,  increasing the grating thickness  as well as the volume ratio of the nanoparticles at the same time even mirrors could be produced. The versatility (species of nanoparticles with distinct properties, volume ratio) as well as the excellent optical quality and reduced shrinkage during preparation \cite{Nakamura-joa09} comprise the major advantages over any other system investigated up to now. Thus the experimental demonstration that nanoparticle-photopolymer composites exhibit a strong PNRE can be regarded as a breakthrough for the future development of versatile neutron diffractive elements and a new generation of cold neutron interferometers.

\begin{acknowledgments}
We are grateful to K. Momose and  S. Yasui for contributing to the sample preparation and Xu Jingjun for discussion. Financial support by the Austrian Science Fund (P-18988, P-20265) and from the Ministry of Education, Culture, Sports, Science and Technology of Japan (grant\# 20360028) is acknowledged. This research has been supported by the European Commission under the 6th Framework Programme through the Key Action: Strengthening the European Research Area, Research Infrastructures. Contract no.: RII3-CT-2003-505925.
\end{acknowledgments}

\end{document}